\documentstyle[12pt,twoside,fleqn,espcrc1,epsfig]{article}
% ...

\newcommand{\AmS}{{\protect\the\textfont2
  A\kern-.1667em\lower.5ex\hbox{M}\kern-.125emS}}
\title{\bf {\large{Experimental Signals of Phase Transition}}}
\author{
M.~D'Agostino\address{{Dipartimento di Fisica and INFN, Bologna,Italy}
\vspace{-.3cm}}
M.~Bruno$^a$, F.~Gulminelli\address{{
LPC Caen (IN2P3-CNRS/ISMRA et Universit\'e), F-14050 Caen
C\'edex, France}\vspace{-.3cm}}, R.~Bougault$^b$, F.~Cannata$^a$,
Ph.~Chomaz\address{{GANIL (DSM-CEA/IN2P3-CNRS)}\vspace{-.3cm}},
F.~Gramegna\address{{INFN Laboratorio Nazionale di Legnaro, Italy}\vspace{-.3cm}},
N.~Le~Neindre\address{{Institut de Physique nucl\'eaire, IN2P3-CNRS, Orsay, France}\vspace{-.3cm}},
A.~Moroni\address{{Dipartimento di Fisica and INFN,Milano, Italy}\vspace{-.3cm}},
G.~Vannini$^a$}

\begin{document}
\maketitle

\vspace{.3cm}
\noindent
{\large \it {8th International Conference on Nucleus-Nucleus
    Collisions, Moscow 2003}}

\begin{abstract}
{The connection between the thermodynamics of charged finite
nuclear systems and the asymptotically measured partitions is
presented. Some open questions, concerning in particular 
equilibrium partitions are discussed. We show a detailed comparison of
the decay patterns in $Au+C, Cu, Au$ central collisions and 
in $Au$ quasi-projectile events. Observation of abnormally large
fluctuations in carefully selected samples of data is reported as 
an indication of a first order phase transition
(negative heat capacity) in the nuclear equation of state.}
\end{abstract}

\section{INTRODUCTION}
In the last 20 years and especially in the most recent ones,
considerable progress has been achieved both theoretically and
experimentally in the investigation of nuclear reaction dynamics and 
thermodynamics in the Fermi energy regime~\cite{chomaz}.

From a theoretical point of view, strong efforts have been devoted to
the understanding of the nuclear equation of state (EoS) either with 
transport theories or statistical approaches. 

From an experimental point of view new generation $4\pi$ detectors
have been developed and are now operating at different accelerator
facilities (Dubna, GANIL, GSI, LNL, LNS, MSU, Texas A-M). 
They are producing a huge amount of exclusive data and new kind 
of analyses. Very rich information has already been extracted from
experimental studies on intermediate energy heavy ion collisions.
Experiments have shown that the final state can be
constrained to select the dynamics of the collision and isolate
events that populate states closely compatible with equilibrium. 
Several investigations have demonstrated that excited 
nuclear systems produced in such collisions undergo in a short 
time scale (100 fm/c) bulk multifragmentation
characterized by final states containing several Intermediate
Mass Fragments (IMF,$Z \ge$ 3). 

A considerable progress has been accomplished on the theoretical as well as 
on the experimental side in order to define and collect a converging 
ensemble of signals connecting multifragmentation to 
the nuclear liquid-gas phase transition~\cite{chomaz} and locating it 
in the phase diagram of nuclear matter~\cite{natowitz}.
The opening of the high fragment multiplicity channel, the onset of  
collective expansion, the flattening of the caloric curves, the fossile
signal of spinodal decomposition, a negative branch of the heat
capacity, the bimodal distribution of exclusive observables and the
finite size and Fisher law scalings have been observed and tentatively 
related to the Equation of State of the nuclear matter.
The possibility of new radioactive beams (RIB) facilities is
now prompting exciting theoretical advances of the isospin
aspects of the EoS, like the density dependence of the 
symmetry energy and the modifications to the spinodal 
instability~\cite{chomaz}.
Experimentally,  first studies on the dependence of the isoscaling
parameters on the isospin of the decaying system~\cite{geraci} already
started exploiting stable beams.
These works seem to indicate an isospin distillation in asymmetric systems.  

All these signals can be considered as circumstantial evidences of a phase
transition, but some of them are still controversial
and need to be further experimentally investigated before 
the phase transition can be definitely assessed.
This work is a contribution to this aim, but only next generation
experiments will allow to reach a comprehensive and detailed
understanding of the phase transition and the nuclear EoS.
\section{THERMODYNAMICS OF NUCLEAR SYSTEMS}
\subsection{EQUILIBRIUM PARTITIONS}
In order to perform thermodynamical analyses, one has to collect
a data sample which corresponds as closely as possible 
to an homogeneous population of the phase space.
Data must be selected such as to isolate a portion of the cross section
where the entire system (or the quasi-projectile) properties 
keep a negligeable memory of the entrance channel dynamics.
This can be experimentally verified checking that 
for a given source the fragmentation pattern  is determined by
the size, charge, energy and average freeze-out volume solely,
independent of the way the source has been formed, e.g. different
impact parameters.
In this case the thermodynamics we can access is a microcanonical
thermodynamics with energy, number of protons and neutrons, 
and average volume as state variables. Indeed the excitation energy can be 
measured on an event-by-event basis by calorimetric techniques.
For any shape of the excitation energy distribution the
events can thus be sorted in constant energy bins, i.e. in
microcanonical ensembles.

In the following we present thermodynamical studies performed on
quasi-projectile events from peripheral 35 A~MeV $Au+Au$
collisions~\cite{michela} and central events from 25 A~MeV
$Au+C$, 25 and 35 A~MeV $Au+Cu$ and 35 A~MeV $Au+Au$
collisions~\cite{fisherexp}, measured at the K1200-NSCL Cyclotron
of the Michigan State University with the MULTICS-MINIBALL
apparatus.

\begin{figure} [hbt]
\vspace{-1.cm}
\begin{center}
\mbox{
\hspace{-.5cm}
\epsfig{file=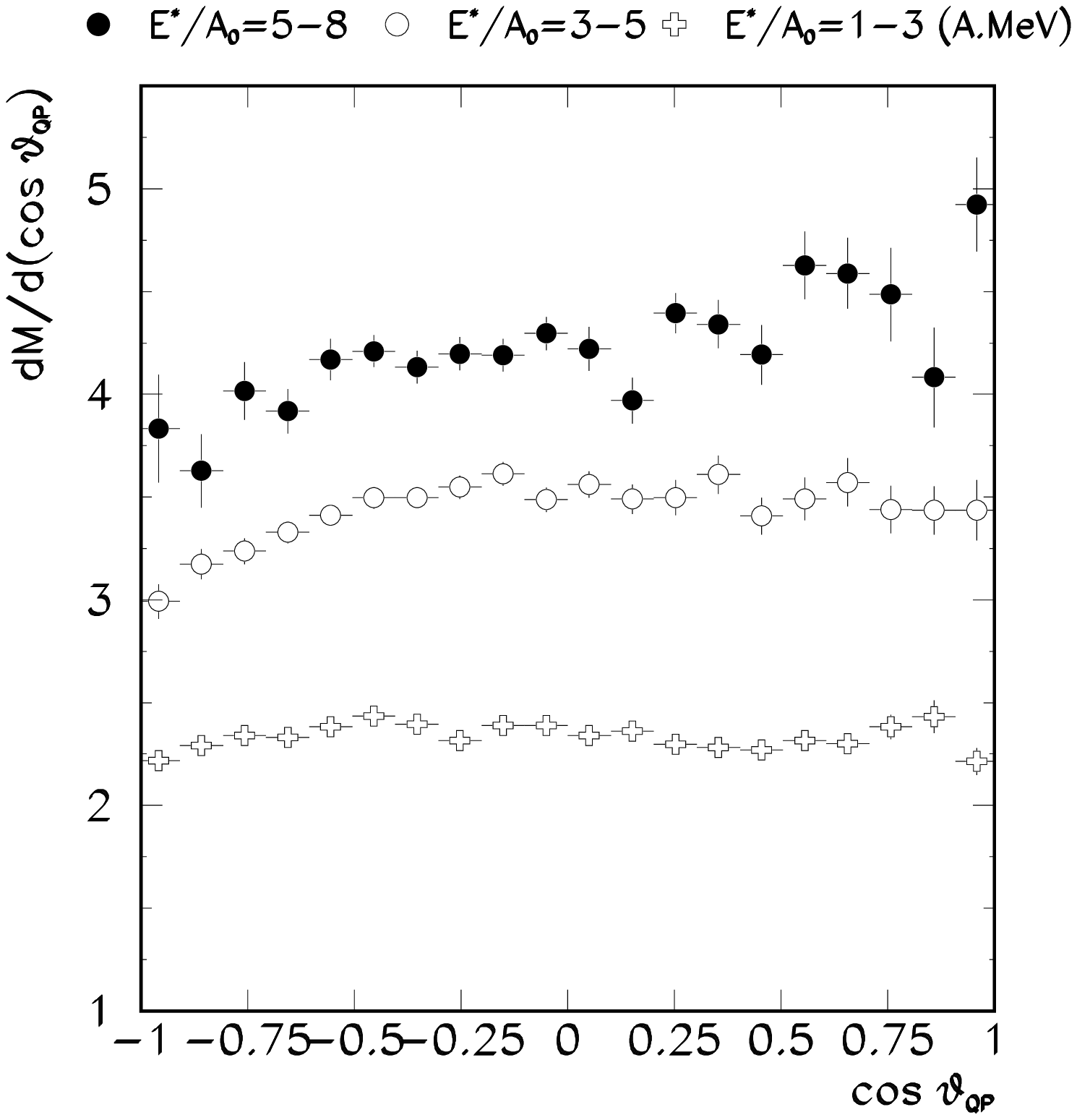,width=6.5cm}
\hspace{1.cm}
\epsfig{file=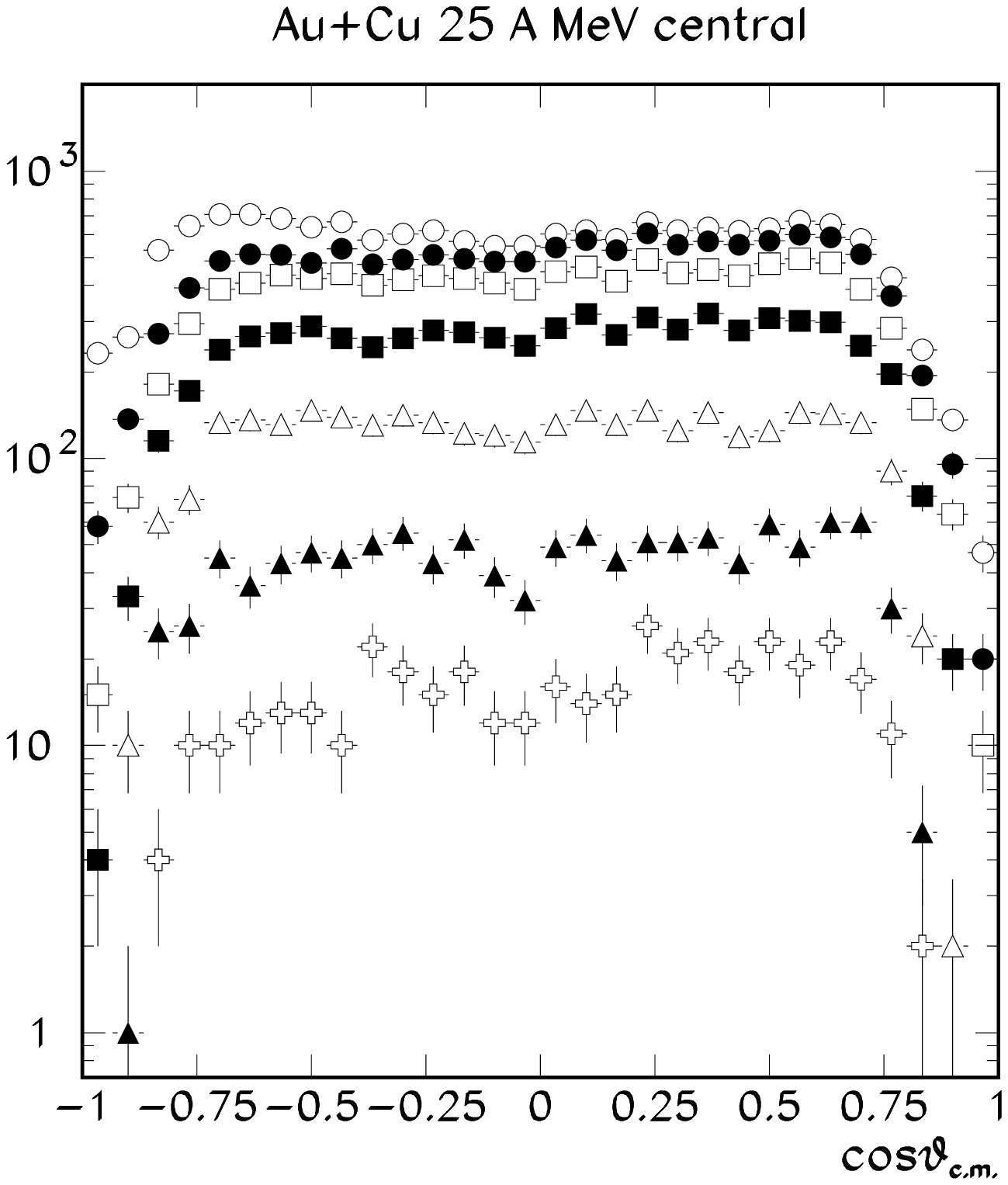,width=6.5cm}
}
\vspace{-1.5cm}
\caption{\label{costcm} \baselineskip=1pt\small
Left: Angular distribution of the fragments ($Z\ge 3$), but the
heaviest one, in the $Au$ quasi-projectile reference frame in different
intervals of the excitation energy.
Right: Angular distribution in the center of mass reference frame for
central $Au+Cu$ events at 25 A~MeV: open circles, full points, open squares 
full squares, open triangles, full triangles, open crosses refer to 
$Z>8$, $18$, $28$, $38$, $48$, $58$ $68$, respectively. 
}
\end{center}
\vspace{-.5cm}
\vspace{-.7cm}
\end{figure}
Single source almost complete events have been selected with a constant
value for the collected charge\footnote{90\% of the total charge for central
collisions and 90\% of the projectile charge for peripheral events}
in each energy bin~\cite{michela,fisherexp}. 
The possible pollution from other
sources has been minimized for central collisions through a shape
analysis and in the case of $Au$ quasi-projectile by substituting
the backward light particle emission by the symmetric of the
forward emission in the quasi-projectile reference frame. The
observed event isotropy (Fig.~\ref{costcm}) indicates that the
directed flow component coming from a memory of the entrance
channel is negligible.

The close similarity between statistical models~\cite{bondorf} and 
data~\cite{michela,fisherexp}, together with the isotropy of fragment 
emission, already suggest that these sets of data are close to a 
statistical equilibrium.
\begin{figure} [bhtp]
\vspace{-.4cm}
\begin{center}
\vspace{-1.4cm}
\epsfig{file=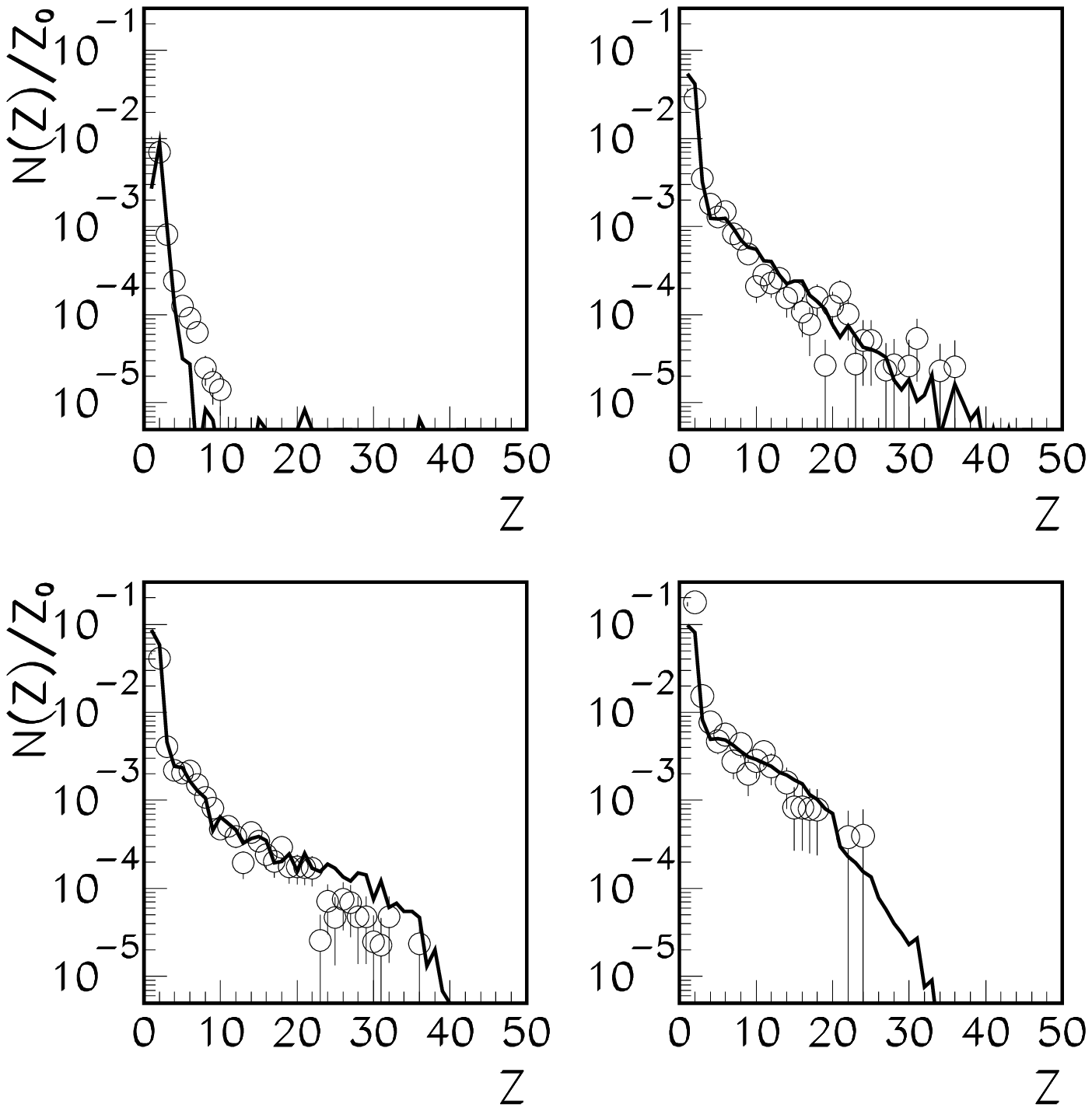,height=8.5cm}
\hspace{-1.5cm}
\epsfig{file=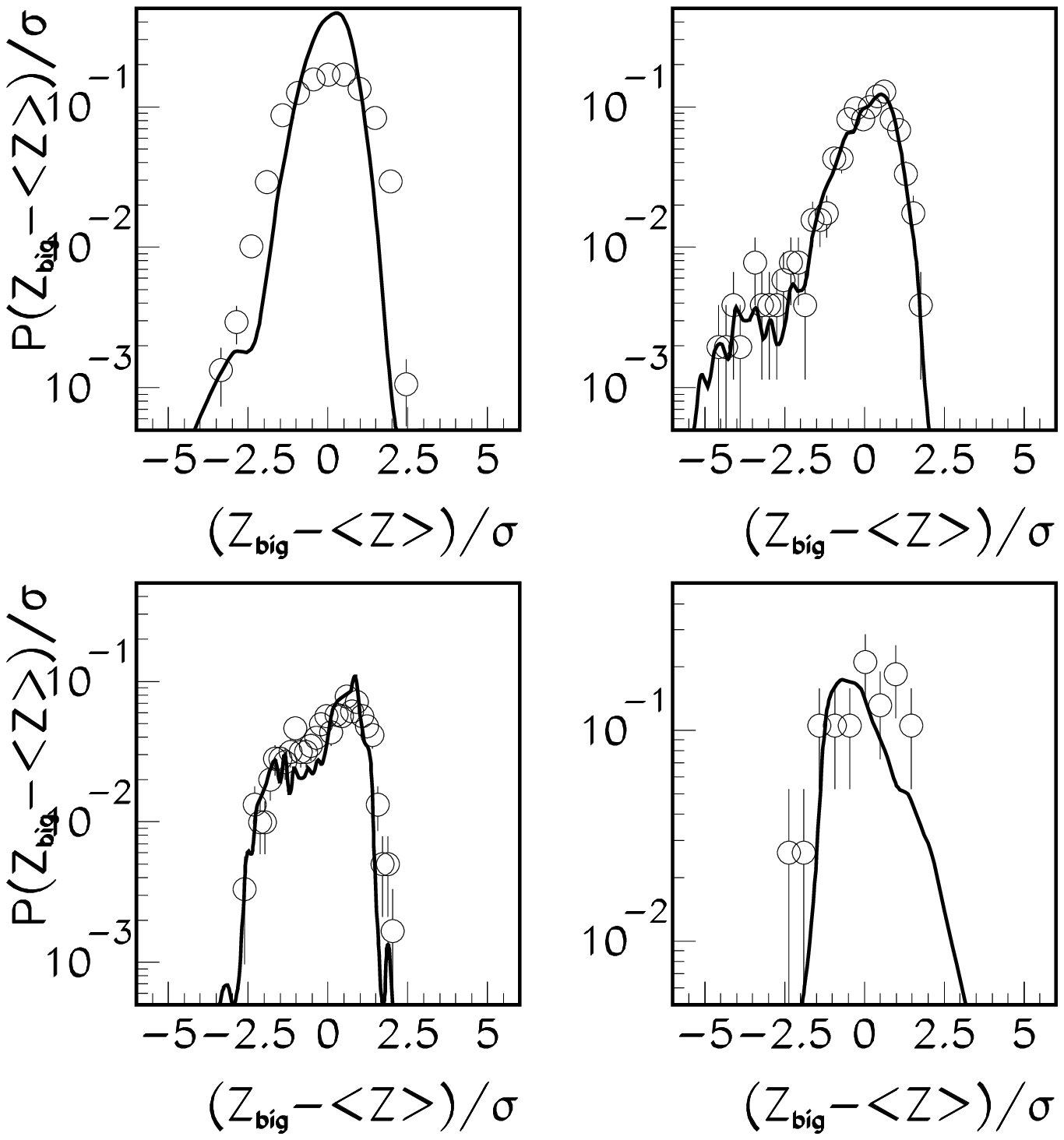,height=8.5cm}
\vspace{-1.5cm}
\end{center}
\vspace{-.3cm}
\caption{\label{3938}\baselineskip=1pt\small
Charge distribution for central collisions (lines) compared 
to the QP distributions (symbols) in the same calorimetric energy
bins. The distributions are  normalized to the charge of the emitting
systems.
Left: all detected fragments, but the heaviest. Right: heaviest fragment.
Fission has been recognized and reconstructed through a correlation
technique\protect{\cite{michela,fisherexp}}.
}
\vspace{-.5cm}
\end{figure}
 
To progress on this point, we can also compare different data sets.
Central events correspond to a narrow distribution of the
excitation energy, while the quasi-projectile data supply a widely
spread excitation function. Therefore it is possible to find for these
data sets common values for the energy deposited in the source and to
compare charge distributions and charge partitions.
In the left four panels of Fig.~\ref{3938} the charge distribution
of all the reaction products, but the largest one, masured for central 
$Au+C$ events at 25 A~MeV, $Au+Cu$ at 25, 35 A~MeV and $Au+Au$ at
35 A~MeV (top left, top right, bottom left, bottom right panels,
respectively) are compared to the quasi-projectile distributions at
the same excitation energy. The calorimetric values of the energy deposited in 
the systems cover a wide range. Indeed their values are 1.6, 3.1, 4.7
and 7.4 A~MeV, respectively. Moreover, the comparison among the charge
distributions of the heaviest fragment in each event (Fig.~\ref{3938})
allows to check the scaling of higher moments like the variances,
even in more detail.

A very good scaling behaviour is also obtained by comparing the
presented sets of data and the charge distributions measured by
the FASA collaboration for the reactions $p, \alpha +Au$
at 8.1, 4 and 14.6 GeV~\cite{fasa} and by the Indra
collaboration for $Xe + Sn$ reaction at 32 A MeV incident energy~\cite{mfra}.

The remarkable scaling between different sets of data means that these 
data samples can be analyzed, at least in a good first approximation, 
within statistical microcanonical methods.
\subsection{PSEUDO-CRITICAL BEHAVIOUR IN NUCLEI}
Since the early 80's, size distributions have been fitted with
power laws~\cite{histo}, and more sophisticated critical
analyses have been performed following theoretical concepts
coming from percolation theory.
More recently, an astonishing good scaling behaviour has been
observed in the EOS data~\cite{eos} and tentatively associated to
the critical point of the nuclear liquid-gas phase transition
expected to occur in nuclear matter in the framework of the Fisher droplet
model~\cite{fisher}.
The debate on the order of the transition has been further
animated by a very recent analysis of the EOS and Isis
data~\cite{eos-isis} which shows a high quality scaling of the
fragment size distribution over a wide range of charges and
deposited energies with an ansatz for the scaling function taken
from the Fisher droplet model. The Fermi gas
"critical" temperature (about 8 MeV) extracted in these papers 
is identified as the temperature of the thermodynamical
critical point and the whole coexistence line of the nuclear
phase diagram is reconstructed under the assumption that the
Fisher model gives a good description of the multifragmentation
phenomenon~\cite{eos-isis}.
In this interpretation, multifragmentation would
correspond to the critical point of the nuclear matter EoS 
(with a lower temperature due to finite size and Coulomb effects),
i.e. to a second order phase transition.

However, the experimental observation of a flattening of the 
caloric curve~\cite{natowitz,caloric} and recent
studies~\cite{karna} in the framework of the SMM
model~\cite{bondorf} (in which the resulting critical temperature is about 20
MeV) point rather to a first order phase transition, 
and this is also suggested by other 
thermodynamical statistical multifragmentation models~\cite{gross}. 
\begin{figure}[hbt]
\vspace{-1.cm}
\mbox{
\vspace{-1.5cm}
\hspace{-.9cm}
\epsfig{figure=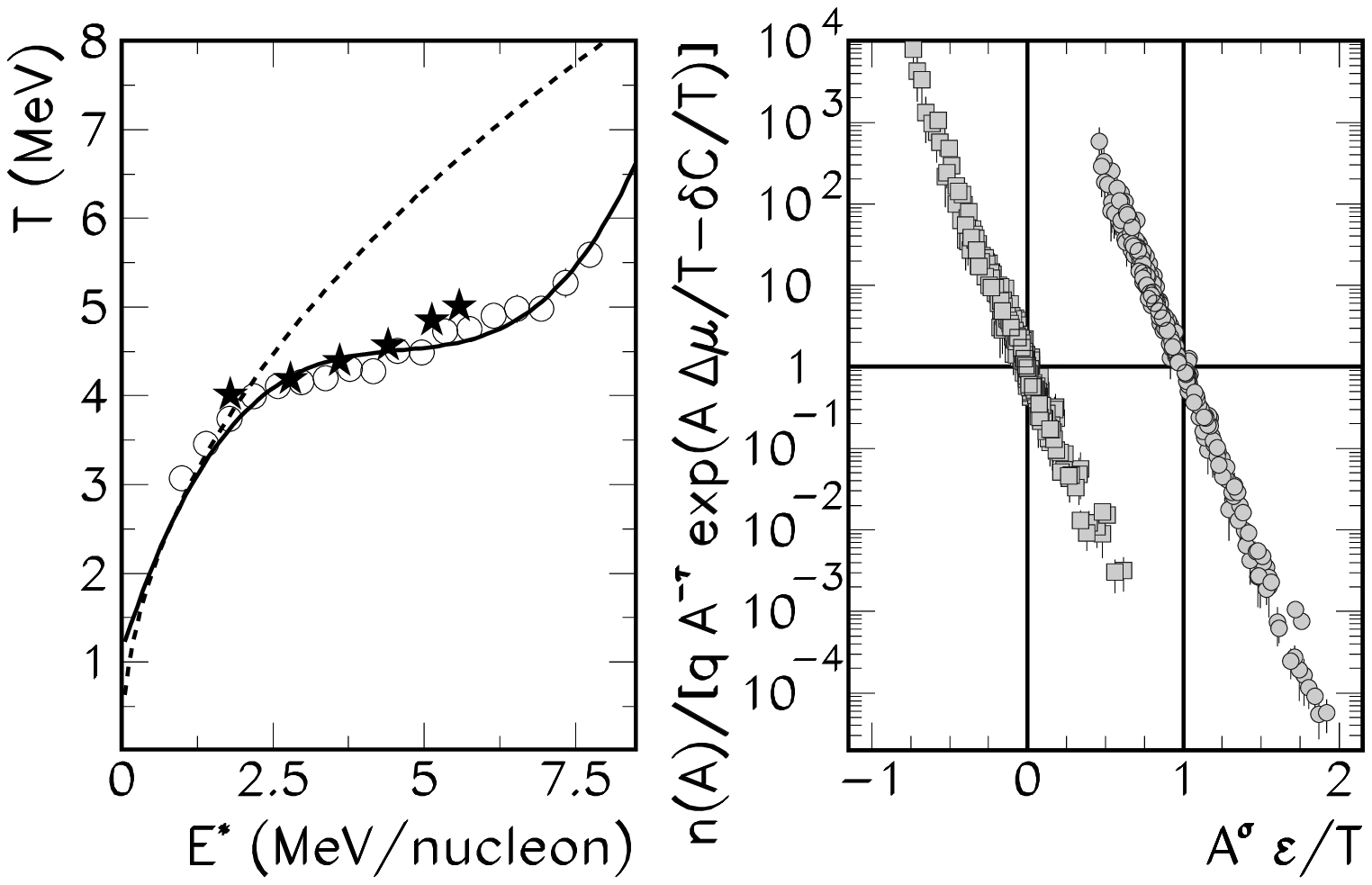,height=6.cm}
\hspace{-1.5cm}
\epsfig{figure=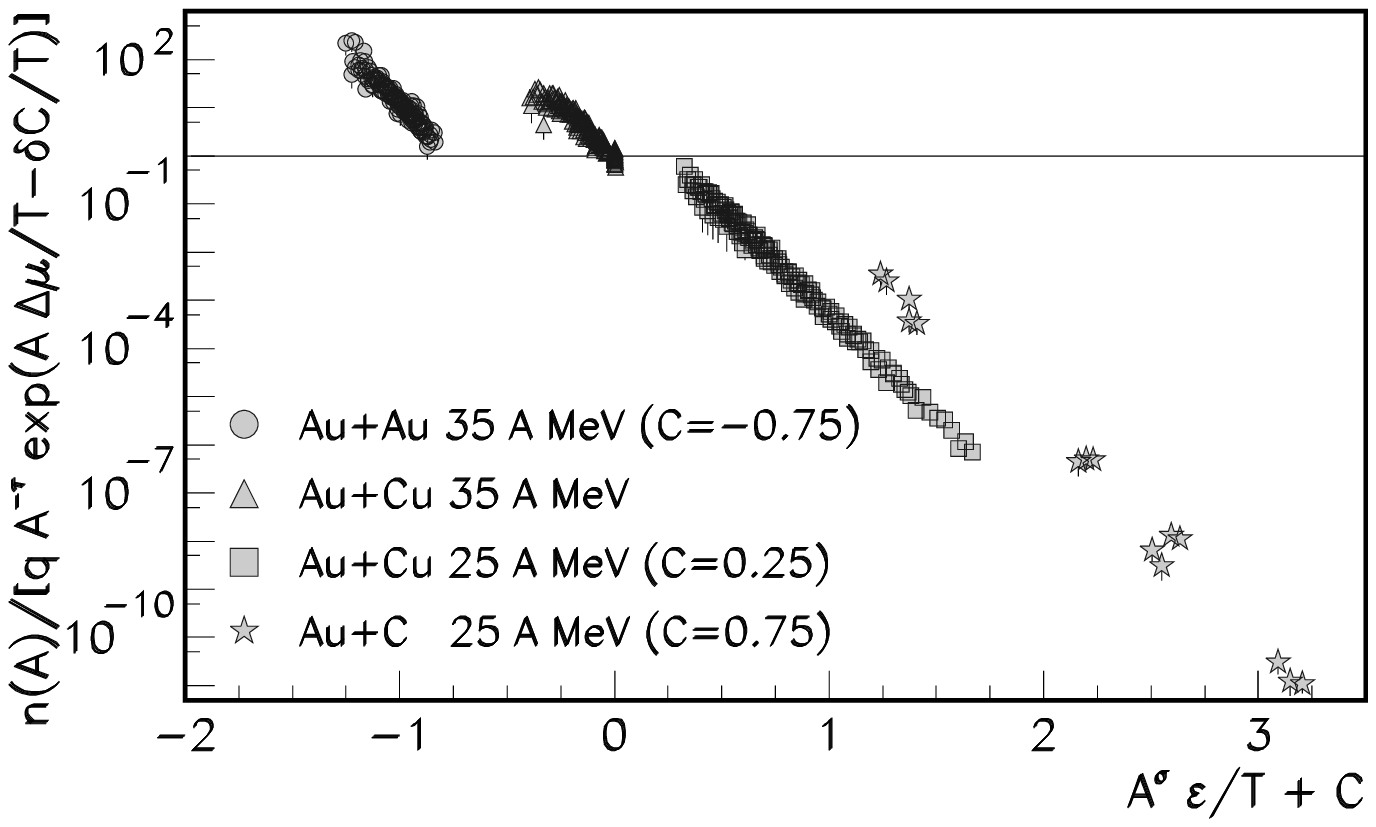,height=6.cm}
}
\vspace{-1.8cm}
\caption{\label{qp}\baselineskip=1pt\small
Left side: different estimations of the QP temperature. 
Circles: kinetic energy thermometer from Eq.(\protect{\ref{temp}}) and
3-rd degree polynomial fit (solid line).
Stars: isotopic thermometer from Ref.\protect{~\cite{pal2002}}.
Dashed line: Fermi gas ansatz $\sqrt{8 E^*}$. Middle panel: QP scaled
yields as a function of the scaled $T$, estimated from 
the solid line (squares) and from the dashed line (circles). 
Right panel: Scaled yields for central collisions. 
In the middle and right panels to represent the results on the same
picture a constant horizontal shift $C$ is applied to the distributions. 
 }
\vspace{-1.2cm}
\end{figure}
One may also wonder what physical meaning can be attributed to 
a Fermi gas estimator of the temperature at excitation energies of 
several A~MeV. Indeed for  the $Au$ quasi-projectile data we are 
discussing, the temperature, evaluated from an isotope and a kinetic 
energy thermometer~\cite{pal2002}, strongly deviates from 
the Fermi gas ansatz (left part of Fig.~\ref{qp}). 

Surprisingly enough, a comparable quality of scaling (Fig.~\ref{qp}) 
and a consistent set of critical exponents can also
be extracted from the $Au$ QP data samples,  for which the
heat capacity evaluation points to a first order phase
transition~\cite{michela,pal2002}, and the quality of the scaling 
does not depend the ansatz adopted for the $T$ parameter. 
The same ansatz for the scaling function applied to central
events~\cite{fisherexp} gives similar exponents and points to the same 
critical-like energy.

However, critical exponents and scale invariance are compatible with 
many different physical phenomena and are not necessarily linked to 
a thermodynamic second order phase transition~\cite{campinew}. 
In particular the observed signals of critical behavior can be
compatible with a first order phase transition, since in different 
statistical models size distributions that mimic a scale invariant 
behavior are observed inside the coexistence zone of small
systems~\cite{prl99}. This means that scaling {\it per s\'e} does not 
demonstrate the existence of a phase
transition, and even less defines its order or allows to localize
the system on the phase diagram. 

On the other hand, the fact that all the analyzed reactions behave as 
a universal multifragmentation process independent of the entrance channel
and directly correlated with the available energy only,
is a strong indication of a microcanonical equilibrium. 
\section{MULTIFRAGMENTATION AND SIGNALS OF PHASE TRANSITION}
It has been shown~\cite{npa} that for a given total
energy the average partial energy stored in a subsystem of the
microcanonical ensemble is a good thermometer while the fluctuations
associated to the partial energy can be used to evaluate the heat
capacity. An example of such a decomposition is given by the
kinetic $E_k$ and the interaction $E_I$ energies.
In particular first order phase transitions are marked by singularities and
negative heat capacities~\cite{gross,npa}, corresponding to
fluctuations anomalously larger than the canonical expectation.
If the system is in statistical equilibrium,
a measurement of anomalous fluctuations at a given energy
is an unambiguous proof of a thermal first order phase transition.

Experimentally, the total energy $E^*$ deposited in the system can be
evaluated event by event by calorimetry. If one is also able to
reasonably estimate for multifragmentation data the relation between 
the measured charge and mass of the reaction products and the value of 
the freeze-out volume, one can obtain the event-by-event Q-value and 
the Coulomb energy, and thus the interaction energy, according to:
\begin{equation}
E_I = \sum_{i=1}^M m_i + E_{coul}^{FO}-m_0  \ \ \ \ \ (i=1,M) \label{e_int}
\end{equation}
where M is the total multiplicity, $m_i$ ($m_0$) are the mass excess of the
primary products (of the source).
The interaction energy fluctuation can then be studied as 
a function of the total energy and the heat capacity can be evaluated
according to:
\begin{equation}
C = \frac{C_k^2 T^2}{C_k T^2 - \sigma_k^2} 
\label{57}
\end{equation}
where $\sigma^2_k=\sigma^2_I$ is the fluctuation of the interaction 
energy $E_I$ from Eq.(\ref{e_int}), $T$ is the temperature, 
and $C_k$ is the kinetic heat capacity that can be evaluated by taking 
the numerical derivative of 
$\langle E_k\rangle=E^* - \langle E_I\rangle$ with respect to $T$.
Eq.(\ref{57}) shows that a negative heat capacity corresponds to partial
energy fluctuations in the microcanonical ensemble that exceed the
corresponding fluctuations in the canonical ensemble ($\sigma_{can}=C_k T^2$). 

\begin{figure} [htb]
\begin{center}
\vspace{-2.5cm}
\mbox{\epsfig{file=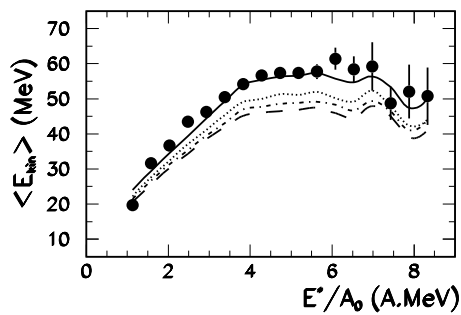,height=6.cm}}
\vspace{-1.5cm}
\end{center}
\vspace{-.5cm}
\caption{\label{vfo}
\baselineskip=1pt\small
Average fragment kinetic energy for QP data. Symbols:
experimental values, lines: many-body Coulomb trajectories for a
volume 2.7 $V_0$ (full line), 4$V_0$ (dotted), 5$V_0$ (dash-dotted),
6$V_0$ (dashed).
}
\vspace{-.7cm}
\end{figure}
\begin{figure} [th]
\begin{center}
\vspace{-2.6cm}
\mbox{\epsfig{file=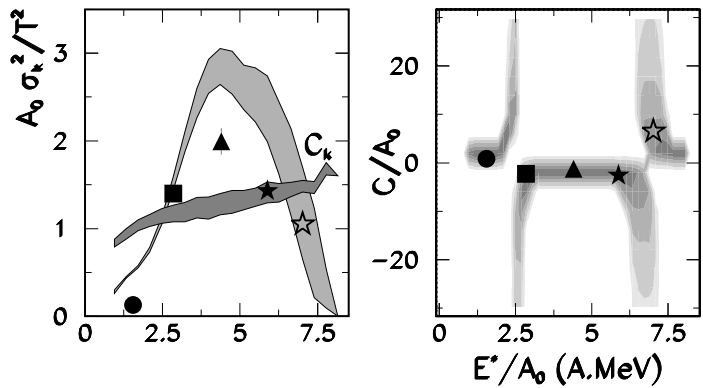,height=7.cm}}
\vspace{-1.5cm}
\end{center}
\vspace{-.5cm}
\caption{\label{cneg} \baselineskip=1pt\small
Left panel: Normalized partial energy fluctuations and
kinetic heat capacity for QP events (grey contours) and
central $Au+C$ (black dots), $Au+Cu$ (squares, triangles), $Au+Au$ reactions
before (open stars) and after subtraction of 1 A~MeV radial flow (black stars).
Left panel: Heat capacity per nucleon of the source for QP events and central
reactions.
}
\vspace{-.8cm}
\end{figure}

The average freeze-out volume can be determined~\cite{pal2002}  from
the detected fragment kinetic energies through a many-body Coulomb trajectory
calculation (Fig.~\ref{vfo}). 
For the sets of data here presented, a volume close to three times the
normal volume reproduces the measured kinetic energies. 
Only for the central $Au+Au$ collisions, the interplay between the
Coulomb energy and the radial flow (about 1 A~MeV) gives an
uncertainty in the determination of the freeze-out volume. Indeed, the 
fragment kinetic energies would also be compatible with zero radial
flow and an increased Coulomb repulsion from a more compact configuration. 
This ambiguity, however, does not affect our main conclusions, as is
evident from the results shown in Fig.\ref{cneg}.

An estimator of the microcanonical temperature $T$ can be obtained by
inverting the kinetic equation of state:
\begin{equation}
\langle E_k\rangle =\big\langle \sum_{i=1}^{M}
\frac{A_i}{a(T)} \big\rangle T^{2}
+\big\langle \frac{3}{2}(M-1)\big\rangle T  \label{temp}
\end{equation}
The unknown parameters of Eq.(\ref{temp})  
are the average side feeding correction $\Delta A$ 
on the fragment masses $A_i$,
and the level density parameter $a$ of primary fragments. 
Since the Coulomb energy is positively correlated with the charged
products multiplicity, the value obtained for the freeze-out volume in
each excitation energy bin depends on the side feeding
correction. This means that $V_{FO}, a_i$ and the percentage of
evaporated particles have to be fixed consistently with an iterative
procedure. 
When this is done~\cite{pal2002}, we find that 
the microcanonical temperature, evaluated through
Eq.(\ref{temp}) results in agreement with the measured isotope
temperature (left panel of Fig.~\ref{qp}).

In order to minimize spurious fluctuations due to unmeasured quantities,
a constant side feeding correction is applied in each excitation energy bin,
and no volume fluctuations are allowed~\cite{pal2002}.
The partial energy fluctuation overcomes the canonical expectation
approximately at the same values of $E^*$ for peripheral and central
collisions data. The same thing is true for central 
$Xe+Sn$ collisions from 32 to 50 A~MeV, measured with the Indra 
device~\cite{pal2002}.
Therefore, the different data sets are fully compatible, the results
do not depend neither on the detector, nor on the data selection.

It is however important to stress that the quantitative study of
nuclear thermodynamics is still at the first stage. 
Only systematic studies of correlated observables, performed with
\begin{figure} [hbt]
\begin{center}
\vspace{-1.5cm}
\mbox{\epsfig{file=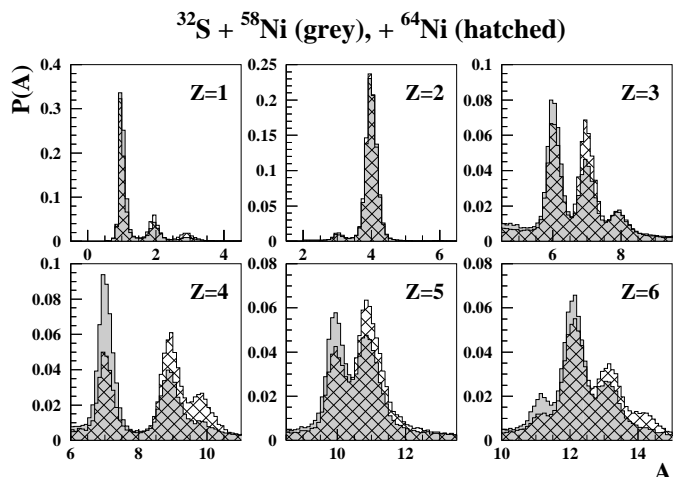,height=7.cm}}
\vspace{-1.5cm}
\end{center}
\vspace{-.5cm}
\caption{\label{compil2} \baselineskip=1pt\small 
Mass probability distribution for a Si-CsI Garfield telescope placed
at $15^o$ in the reactions $^{32}$S+$^{58}$Ni (grey histograms) and 
$^{32}$S+$^{64}$Ni (hatched histograms) at 15 A MeV.
}
\end{figure}
sophisticated experimental devices, with low energy threshold, high
granularity and isotopic resolution will bring a detailed
understanding of the phase transition and will allow to locate the 
position of the observed multifragmentation in the phase diagram of nuclear 
matter~\cite{chomaz,natowitz}.
Recent investigations on the opening of the multifragmentation
channel, on the caloric curve and on the limiting
temperature for $A \approx 100$ systems started at moderate incident energies
($\sim$ 15 A~MeV) at INFN Laboratori Nazionali di Legnaro with the
Garfield apparatus. First results~\cite{fizika} indicate that already
at an excitation energy about 3 A~MeV, multifragmentation events with 
nearly equal size fragments represent a non negligible fraction of
the decay channels. The very good mass resolution of the Garfield Si-CsI
telescopes (Fig.\ref{compil2}) allows an isotope analysis.
The dependence on the isospin of the entrance channel is apparent.
Isoscaling analyses, now in progress, will bring
information on the isospin distillation at the onset of the
multifragmentation~\cite{geraci}. 

\small
\vspace{.2cm}
\noindent
The authors would like to thank Garfield collaboration for kindly 
providing preliminary results.

\end{document}